\documentstyle[twocolumn]{jpsj}
\input{epsf.tex}

\title{Exact Ground States in Spin Systems with Orbital 
Degeneracy}
\author{Kazuhito \sc{Itoh}
\footnote{E-mail address: kzitoh@taka2.issp.u-tokyo.ac.jp}}
\inst{Institute for Solid State Physics, University of 
Tokyo, Roppongi, Minato-ku, Tokyo 106-8666}
\kword{orbital degeneracy, spin systems, valence bond state, 
orbital 
order}
\recdate{ }
\abst{
     We present exact ground states in spin models 
  with orbital degeneracy in one and higher dimensions. 
  A method to obtain the exact ground states of the models
  when the Hamiltonians are composed of the products 
  of two commutable operators is proposed. 
  For the case of the spin-1/2 model with two-fold degeneracy 
  some exact ground states are given, such as the Valence-Bond 
(VB), 
  the magnetically ordered, and the orbitally ordered states 
  under particular parameter regimes. 
  We also find the models with the higher spin and degeneracy  
  which have the new types of VB ground states in the spin and 
the 
orbital 
  sectors.    
}

\begin{document}
\sloppy
\maketitle 

 Quantum spin systems with orbital degeneracy have attracted 
our 
theoretical and experimental interests
\cite{Kugel,Inagaki,Castellani,Feiner,Li,Yamamura,Pati,Santro,
Arovas,Ohkawa,SU4}. 
These systems have rich phases such as the spin order, the 
orbital 
order, 
the spin liquid, and the orbital liquid. 
Examples of these systems are the Mott insulator of 
the transition metal (Mn, Cu, Cr, etc.) 
oxides\cite{Inagaki,Castellani},  
the dynamical Jahn-Teller molecular systems with a phonon 
coupling
(TDAE-${\rm C}_{60}$)\cite{Santro,Arovas},
and the heavy fermion systems in the insulator 
phase\cite{Ohkawa}.
The strong interplay between the orbital and the magnetic 
ordering 
is 
interesting. 
Other interest is the existence of the spin and the orbital 
disorder 
states 
which are induced by the frustrations with the large quantum 
fluctuation
arising from the spin-orbital interactions. 
There might already exist a physical realization of a quantum 
spin-orbital 
liquid in three dimension: ${\rm LiNiO}_2$\cite{Yamamura}. 
The simplest model of the electron system with orbital 
degeneracy is 
the two-fold degenerate Hubbard Hamiltonian with Hund's rule 
coupling. 
In the strong coupling limit, the charge excitation has a 
large gap at 1/4-filling and this Hamiltonian is an effective 
model 
in the insulator phase such as the above spin systems.  
 
  In the present letter, we consider the spin-orbital models 
and 
give the exact ground states for the spin-1/2 and two-fold 
degenerate 
case. 
We extend the model to more than three-fold degenerate and 
half and 
higher spin cases which facilitate the higher dimensional VB 
states 
under 
certain parameter regimes. 
These VB states are the new types of the VB ground states and 
kinds 
of 
the Resonating Valence Bond (RVB) states. 
   
First, we consider the Hamiltonian is defined by 
\begin{eqnarray}
{\cal H} = \sum_{(i,j)} 
({\cal U}_{i,j} - \lambda_{0}^{{\rm u}})
({\cal V}_{i,j} - \lambda_{0}^{{\rm v}}),
\end{eqnarray} 
where the sum runs over all nearest-neighbor bonds $(i,j)$. 
${\cal U}_{i,j}$ and ${\cal V}_{i,j}$ 
denote some Hermitian local operators whose lowest eigenvalues 
are 
given 
by 
$\lambda_{0}^{{\rm u}}$ and $\lambda_{0}^{{\rm v}}$, 
respectively. 
If ${\cal U}_{i,j}$ and ${\cal V}_{i,j}$ commute with each 
other: 
$[{\cal U}_{i,j},{\cal V}_{i,j}]=0$, 
the eigenstate of ${\cal H}$ is described by 
$|\Psi\rangle = |{\rm u}\rangle \otimes |{\rm v}\rangle$, 
where $|{\rm u(v)}\rangle$ denotes a state vector in u(v) 
subspace. 
The local Hamiltonian is positive semi-definite, 
and hence the total Hamiltonian is also positive 
semi-definite. 
The following two types of states are the possible ground 
states of 
the 
Hamiltonian. 

{\it Type(i)}.---One possible ground state is written by
$|\Psi_{{\rm G}}(A,B)\rangle 
= |\Phi^{\rm u}(A)\rangle \otimes |\Phi^{\rm v}(B)\rangle$.
 $|\Phi^{{\rm u}}(A)\rangle$ and $|\Phi^{{\rm v}}(B)\rangle$ 
 are given by 
\begin{eqnarray}
|\Phi^{{\rm u}}(A)\rangle &=& \prod_{(i,j) \in A} 
|i,j\rangle_{0}^{{\rm u}},
\quad 
|\Phi^{{\rm v}}(B)\rangle = \prod_{(i,j) \in B} 
|i,j\rangle_{0}^{{\rm 
v}},
\end{eqnarray}
where $|i,j\rangle_{0}^{{\rm u(v)}} = \sum_{n} 
C_{n}^{i,j}|\phi_{0}^{(n)} 
\rangle_{i,j}^{{\rm u(v)}}(n=1,...,N)$ and 
$|\phi_{0}^{(n)} \rangle_{i,j}^{{\rm u(v)}}$ is the lowest 
eigenstate 
of the local operator ${\cal U}({\cal V})$ on sites $i,j$, 
whose degeneracy is $N$. 
$A$ and $B$ denote {\it a bond-covering} (a configuration of 
the 
bonds) 
in u and v subspaces, respectively, 
which are constrained by $A \cup B = \{$ {\it all bonds }$\}$. 
One site has $z_{\rm u}$ and $z_{\rm v}$ bonds 
on the bond-covering $A$ and $B$, respectively, 
obeying $z \le z_{\rm u} + z_{\rm v}$, 
where $z$ denotes a lattice coordination number.
Obviously, ${\cal H}|\Psi_{{\rm G}}(A,B)\rangle = 0$, thus 
$|\Psi_{{\rm G}}(A,B)\rangle$ is the ground state, 
because the eigenvalues of the Hamiltonian are positive 
semi-definite and 
$|\Psi_{{\rm G}}(A,B)\rangle$ is the lowest eigenstate with an 
eigenvalue of zero. 
The ground-state degeneracy depends on the number of the 
bond-coverings.  

{\it Type(ii)}.---If the state $|\Phi^{\rm u(v)}\rangle$ in 
one 
sector obeys 
${\cal H}|\Phi^{\rm u(v)}\rangle = 0$, where 
$|\Phi^{\rm u(v)}\rangle = \prod_{(ij)} |i,j\rangle_{0}^{{\rm 
u(v)}}$, 
the ground state of the global Hamiltonian 
is given by 
$|\Psi_{{\rm G}}\rangle = |\Phi^{\rm u(v)}\rangle \otimes 
|{\rm v(u)}\rangle$, where $|{\rm v(u)}\rangle$ denotes 
an arbitrary state in v(u) space. 
Because ${\cal H}$ is positive semi-definite and ${\cal 
H}|\Psi_{{\rm 
G}}\rangle = 0$, and hence $|\Psi_{{\rm G}}\rangle$ is the 
lowest 
eigenstate with an eigenvalue of zero. 
In addition, the ground state of the Hamiltonian 
${\cal H} \to {\cal H} + \alpha \tilde{{\cal H}}_{{\rm v(u)}}$ 
$(\alpha \ge 0)$
is given by 
$|\Psi_{{\rm G}}\rangle = |\Phi^{\rm u(v)}\rangle \otimes 
|\tilde{\Phi}^{\rm v(u)}\rangle$, where $|\tilde{\Phi}^{\rm 
v(u)}\rangle$ 
is the ground state of the $\tilde{{\cal H}}_{{\rm v(u)}}$ 
with the 
ground-state energy $\tilde{E}_{0}^{{\rm v(u)}}$, 
since ${\cal H} - \alpha\tilde{E}_{0}^{{\rm v(u)}}$ is 
positive 
semi-definite and $({\cal H}- \alpha\tilde{E}_{0}^{{\rm 
v(u)}})|\Psi_{{\rm 
G}}\rangle = 0$, and hence $|\Psi_{{\rm G}}\rangle$ is the 
lowest 
eigenstate with the eigenvalue $\alpha\tilde{E}_{0}^{{\rm 
v(u)}}$. 
 
Moreover, the ground states of the Hamiltonian: 
${\cal H} = \sum_{(i,j)} a_{n,m} ({\cal U}_{i,j} - 
\lambda_{0}^{{\rm 
u}})^{n} 
({\cal V}_{i,j} - \lambda_{0}^{{\rm v}})^{m}$ $(a_{n,m} \ge 0, 
n,m 
\ge 
1)$ 
are also the same as those of the two above cases.
     
 Using the above results we discuss the ground states of 
the spin-orbital Hamiltonian: 
\begin{eqnarray}
{\cal H} &=& \sum_{(i,j)} J_{\rm s} {\bf S}_{i} \cdot {\bf 
S}_{j} +
J_{{\rm \tau}} (\tau_{i}^{x} \tau_{j}^{x} 
+\tau_{i}^{y} \tau_{j}^{y} 
+ \Delta \tau_{i}^{z} \tau_{j}^{z} ) \nonumber \\
& & {} + J_{{\rm s\tau}} ({\bf S}_{i} \cdot {\bf S}_{j} )
(\tau_{i}^{x} \tau_{j}^{x} 
+\tau_{i}^{y} \tau_{j}^{y} 
+ \Delta^{\prime} \tau_{i}^{z} \tau_{j}^{z}), \label{eqn:Ham1}
\end{eqnarray}
where ${\bf S}_{i}$ denotes the spin operator and 
$\tau_{i}^{a}$ $(a=x,y,z)$ denotes 
the $a$-component of the pseudo-spin operator. 
The pseudo-spin $+$ and $-$ ($\tau_{i}^{z} = 1/2$ and $-1/2$)
correspond to the doubly degenerate orbital levels. 
This Hamiltonian is derived from the two-fold degenerate 
Hubbard 
model 
with Hund's rule coupling in the strong coupling limit 
at 1/4-filling by the second-order perturbation\cite{Kugel}. 
The Hamiltonian can be rescaled by $|J_{{\rm s\tau}}|$ and let 
us 
set 
$|J_{{\rm s\tau}}| = 1$, 
${\cal U}_{i,j} = J_{{\rm s\tau}} {\bf S}_{i} \cdot {\bf 
S}_{j}$, 
and 
${\cal V}_{i,j} = \tau_{i}^{x} \tau_{j}^{x} + \tau_{i}^{y} 
\tau_{j}^{y} 
+ \Delta^{\prime} \tau_{i}^{z} \tau_{j}^{z}$. 
The lowest eigenstate(s) of ${\cal V}_{i,j}$ is 
$|{\rm singlet} \rangle_{i,j}^{\rm \tau} = (|+ - \rangle_{i,j} 
- | - 
+ 
\rangle_{i,j})/\sqrt{2}$   with the eigenvalue 
$-\frac{\Delta^{\prime}}{4} - \frac{1}{2}$ 
for $\Delta^{\prime} > -1$, and 
are $|+ + \rangle_{i,j}$ and $|- - \rangle_{i,j}$
with the eigenvalue $\frac{\Delta^{\prime}}{4}$ 
for $\Delta^{\prime} \le -1$. 
For $J_{{\rm s \tau}} = 1$ the lowest eigenstate of ${\cal 
U}_{i,j}$ 
is 
singlet: $|{\rm singlet} \rangle_{i,j}^{\rm s} = (| \uparrow 
\downarrow 
\rangle_{i,j} - | \downarrow \uparrow 
\rangle_{i,j})/\sqrt{2}$, with 
the eigenvalue 
$-\frac{3}{4}$. 
We can find the following exact ground states in one and 
higher 
dimensions. 

 {\it Dimerized ground states in one dimension.}---For 
$J_{\rm s} = \frac{\Delta^{\prime}}{4} + \frac{1}{2}$ 
($\Delta^{\prime} = \Delta \ge -1$) and $J_{{\rm \tau}} = 
\frac{3}{4}$, 
the Hamiltonian (\ref{eqn:Ham1}) can be written as 
${\cal H} = \sum_{(i,j)} 
({\cal U}_{i,j} + \frac{3}{4})
({\cal V}_{i,j} + \frac{\Delta^{\prime}}{4} + \frac{1}{2}) 
+$constant. 
The ground states are {\it type (i)} and given by 
\begin{eqnarray}
   |\Psi_{\rm G} \rangle = \prod_{i,j} 
   |{\rm singlet} \rangle_{i,i+1}^{\rm s} \otimes 
   |{\rm singlet} \rangle_{j,j+1}^{\rm \tau},
\end{eqnarray} 
where $i=$even(odd) and $j=$odd(even). 
The ground states are the products of nearest-neighbor singlet 
dimers and 
two-fold degenerate. 
The translational symmetry is spontaneously broken 
although the interaction is only nearest-neighbor  
and no bond-alternation. 
The ground-state energy is 
  $E_0 = -\frac{3}{16}(\Delta+2)L$, where $L$ denotes the 
number of 
sites.
At $\Delta = 1$, Kolezhuk and Mikeska\cite{Kolezhuk} have 
already 
given 
the exact ground states in the ladder model with a leg-leg 
biquadratic interaction by the matrix product ansatz. 
In the limit $\Delta \to \infty$, the lowest states of the 
local 
Hamiltonian 
in the orbital part are two-fold degenerate 
($|+ - \rangle_{j,j+1}$ and $|- + \rangle_{j,j+1}$). 
The ground states of the global Hamiltonian in the orbital 
space are 
also 
two-fold degenerate and the orbital order 
(the N\'eel order in the pseudo-spin part) exists. 
At this point the system has a first-order phase transition. 
In other words, if an infinitesimal xy-element exists in the 
pseudo-spin 
space, 
then the ground states are the singlet, disordered states. 
The interaction of the transverse direction between spins and 
orbitals drastically enhances the 
quantum fluctuation.      

  {\it Ferromagnetic ground state in the pseudo-spin 
sector}---
When the Hamiltonian (\ref{eqn:Ham1}) can be expressed as 
  \begin{eqnarray}
{\cal H} &=& \sum_{(i,j)} 
({\cal U}_{i,j} + \frac{3}{4})
({\cal V}_{i,j} - \frac{\Delta^{\prime}}{4})
+ \alpha{\bf S}_{i} \cdot {\bf S}_{j}
 \nonumber \\
& &  { } + \beta{\cal H}_{{\rm XXZ}}^{{\rm \tau}}(\gamma) 
 + {\rm constant}, \nonumber 
 \end{eqnarray} 
where  
$\alpha = J_{{\rm s}}+\frac{\Delta^{\prime}}{4}$ 
($\Delta^{\prime} 
\le 
-1$), 
$\beta = J_{{\rm \tau}} - \frac{3}{4} \ge 0$, and $\gamma \le 
-1$ 
and  
${\cal H}_{{\rm XXZ}}^{\tau}(\gamma)$ 
denotes the Hamiltonian of the XXZ model in the pseudo-spin 
space 
and is defined by 
  \begin{eqnarray} 
   {\cal H}_{{\rm XXZ}}^{\rm \tau} (\gamma) = \sum_{(i,j)} ( 
\tau_{i}^{x} 
\tau_{j}^{x} 
       + \tau_{i}^{y} \tau_{j}^{y} + \gamma \tau_{i}^{z} 
   \tau_{j}^{z} ), 
   \end{eqnarray}
namely, for $J_{\rm \tau} \ge \frac{3}{4}$, 
$J_{{\rm s\tau}}=1$, and 
$\Delta  = \frac{3}{4J_{\rm \tau}}(\Delta^{\prime} - \gamma) + 
\gamma$
($\Delta^{\prime} \le -1$), 
the ground state is {\it type (ii)} and is a ferromagnetic 
state
in the pseudo-spin space.
In the spin space the ground state is equivalent 
to that of the antiferromagnetic (AF) Heisenberg model 
for $J_{\rm s} > - \Delta^{\prime}/4$ and 
of the ferromagnetic Heisenberg model 
for $J_{\rm s} < - \Delta^{\prime}/4$. 
For the ferromagnetic spin states  
the ground-state energy is given by 
$E_0 = -\frac{1}{4} \left[ J_{\rm s} + J_{\rm \tau}\gamma + 
\Delta^{\prime} - \frac{3}{4}\gamma \right] L $ 
in any dimension.
    In $D=1$ the AF Heisenberg model is solved by the Bethe 
Ansatz and has unique, massless singlet ground 
state\cite{Hulthen}.
The one-dimensional ground-state energy for $J_s > - 
\Delta^{\prime}/4$ is 
given by 
$E_0 = -\frac{1}{4} \left[ J_{\rm s}(4\ln 2-1) + J_{{\rm 
\tau}}\gamma+ 
\Delta^{\prime}(\ln 2 +\frac{1}{2}) - \frac{3}{4}\gamma 
\right] L $. 
For $D >2$ the existence of an antiferromagnetic long range 
order(AFLRO) 
in the AF Heisenberg model has been rigorously 
proved\cite{KLS}, 
but in $D=2$ there is no proof of the existence of the AFLRO. 
Adding an interaction term 
$\delta\sum_{(i,j)} S_i^z S_{j}^z$ ($\delta > 0$) 
to the Hamiltonian
at $J_{\rm s} = -\frac{\Delta^{\prime}}{4}$ and $J_{\rm \tau} 
= 
\frac{3}{4}$, 
the ground states in the spin sector are equivalent to those 
of 
the AF Ising model. 
Thus, the system has the AFLRO on the bipartite lattices 
in arbitrary dimensions with infinitesimal anisotropy of the 
$z$-direction. 
On the other hand, 
at $J_{\rm s} = -\frac{\Delta^{\prime}}{4}$ and $J_{\rm \tau} 
= 
\frac{3}{4}$ 
the spin liquid 
states are induced by the infinitesimal perturbations such
as an inhomogeneous interaction or a frustration
reflecting a lattice topology. 
For example, let us consider the 1/5-depleted square lattice, 
which has the same topology as the ${\rm CaV}_4{\rm O}_9$ 
lattice\cite{Ueda}, 
and consists of the four-site squares 
and the bridge bonds among the squares.
If one adds an interaction term
$J_1 \sum_{(i,j) \in {\rm squares}} 
{\bf S}_{i} \cdot {\bf S}_{j}$ $(J_1 > 0)$ 
to the Hamiltonian,
the exact ground state is a product of the four-spin singlet 
plaquette 
RVB (PRVB) states on the squares. 
On the other hand, adding an interaction term 
$J_2 \sum_{(i,j) \in {\rm bridges}} 
{\bf S}_{i} \cdot {\bf S}_{j}$ $(J_2 > 0)$
 to the Hamiltonian, 
 the exact ground state is a product of the singlet dimer 
states on 
the bridges. 
We can exactly construct the PRVB and the dimer states 
without the next-nearest-neighbor interaction.
An orbital long range order (ORLO) makes the systems have the 
spin 
liquid state under certain parameter regions.   

 {\it Ferromagnetic ground state in the spin sector }---
Next we set $J_{{\rm s\tau }} = -1$ and the lowest eigenstates 
of 
${\cal U}_{i,j}$ are triplet whose eigenvalue is 
$-\frac{1}{4}$. 
When the Hamiltonian (\ref{eqn:Ham1}) can be expressed as  
  \begin{eqnarray}
{\cal H} &=& \sum_{(i,j)} 
({\cal U}_{i,j} + \frac{1}{4})
({\cal V}_{i,j} +\frac{\Delta^{\prime}}{4}+\frac{1}{2})
+ \alpha{\bf S}_{i} \cdot {\bf S}_{j} \nonumber \\
& & { } +\beta{\cal H}_{{\rm XXZ}}^{\tau}(\gamma)
 + {\rm constant}, \nonumber 
 \end{eqnarray}
 where $\alpha=J_{\rm s} 
+\frac{\Delta^{\prime}}{4}+\frac{1}{2} 
 \le 0$ ($\Delta^{\prime} \ge -1$) 
 and $\beta = J_{\rm \tau}-\frac{1}{4} \ge 0$, 
namely, for $J_{\rm s} \le - \frac{\Delta^{\prime}}{4} - 
\frac{1}{2}$, 
$J_{\rm \tau} \ge  \frac{1}{4}$, and 
$\Delta = \frac{\Delta^{\prime} - \gamma}{4J_{\rm \tau}} 
+\gamma$
($\Delta^{\prime} \ge -1$), 
the ground state is {\it type(ii)} and 
a ferromagnetic state in the spin sector. 
In the orbital sector the ground state is equivalent to that 
of the 
XXZ 
Hamiltonian ${\cal H}_{{\rm XXZ}}^{\tau}(\gamma)$.
On a square lattice the AF XXZ model has the AFLRO 
for $\gamma > 1.66$ 
that has been proved by Kubo and Kishi and Nishimori {\it 
et.al.}, 
using the infrared bounds method\cite{Kubo-Kishi,Nishimori}. 
This AFLRO corresponds to an OLRO in the orbital space. 
The state with the OLRO is what the electrons alternately 
occupy two 
orbital 
levels. 
   
 Next, we consider the model with the spin-$\frac{1}{2}$ 
and the pseudo-spin-1 in one and higher dimensions described 
by
   \begin{eqnarray} 
   {\cal H} = \sum_{(i,j)} [{\bf S}_{i} \cdot {\bf S}_{j} +        
    \alpha ][ {\bf \tau}_{i} \cdot {\bf \tau}_{j} 
   + \beta ({\bf \tau}_{i} \cdot {\bf \tau}_{j})^2 + \gamma ].
   \end{eqnarray}      
This model is regarded as the effective model with the 
spin--$\frac{1}{2}$ 
for the 3-fold orbital degeneracy.
For simplicity, we discuss only the isotropic case. 

   In one dimension for $\alpha = 3/4$, 
$0 \le \beta < 1/3$, and $\gamma = 2-4\beta$ the exact ground 
states 
are type(i) and
the singlet dimer states in both the spin and the orbital 
spaces,  
because one can write the Hamiltonian as 
  ${\cal H} = \sum_{(i,j), n} a_n 
  [{\bf S}_{i} \cdot {\bf S}_{j} + 3/4 ]
  [ {\bf \tau}_{i} \cdot {\bf \tau}_{j} + 2 ]^{n}$,
where $a_n \ge 0$ and $n > 0$. 
These states are two-fold degenerate 
and the translational symmetry is spontaneously broken.
For the case of $\alpha = 3/4$ and $\beta = 1/3$, 
the orbital ground state is equivalent to a Valence Bond Solid 
(VBS) 
state
of the Affleck-Kennedy-Lieb-Tasaki model\cite{AKLT}.
In the spin sector the ground state is equivalent to 
that of the AF Heisenberg model for $\gamma > 2/3$
and of the ferromagnetic Heisenberg model for $\gamma < 2/3$ 
(type(ii)).
At $\beta = 1/3$ and $\gamma = 2/3$ in the orbital sector the 
VBS 
state 
and  
the dimerized states are degenerate 
and in the spin sector all states are degenerate, 
and hence this point is multi-critical. 
     
    At $(\alpha,\beta,\gamma 
)=(\frac{3}{4},\frac{1}{3},\frac{2}{3})$ 
the Hamiltonian is reduced to 
    ${\cal H} = 2 \sum_{(i,j)} 
    {\cal P}_{i,j}^{(S=1)}{\cal P}_{i,j}^{(\tau=2)}$.
Here ${\cal P}_{i,j}^{(S=1 )}$ and ${\cal P}_{i,j}^{(\tau=2)}$
are the projection operators onto the subspace of the total 
spin 
$S=1$ 
and the total pseudo-spin $\tau = 2$ on the two sites $i,j$, 
respectively. 
In arbitrary bipartite lattices with the coordination number 
three, 
e.g., a 2-leg ladder, a hexagonal lattice, a 1/5-depleted 
square 
lattice, 
we can find that the ground states are type(i) and given by
    \begin{eqnarray} 
    |\Psi_{\rm G} (A,B) \rangle &=& | {\rm Dimer} \rangle^{\rm 
s} 
\otimes 
     |{\rm VB}\rangle^{{\rm \tau}},
     \\
    | {\rm Dimer} \rangle^{\rm s}  &=& \prod _{(i,j) \in A} 
    ( a_i^{\dagger} b_j^{\dagger} - b_i^{\dagger} 
a_j^{\dagger} ) | 
0 
\rangle, 
    \nonumber \\
    | {\rm VB} \rangle^{{\rm \tau}} &=& \prod _{(i,j) \in B} 
     ( c_i^{\dagger} d_j^{\dagger} - d_i^{\dagger} 
c_j^{\dagger} ) | 
0 
\rangle,
     \nonumber   
    \end{eqnarray} 
where $a_i^{\dagger}(b_i^{\dagger})$ and 
$c_i^{\dagger}(d_i^{\dagger})$, 
denote the Schwinger boson's creation operators 
that create the $S=1/2$ up(down) spin  and the $\tau = 1/2$ 
$+(-)$ 
pseudo-spin at the site $i$, respectively. 
A and B denote a configuration of dimers ($z_s=1$) 
and a configuration of VB's ($z_\tau=2$) which has two bonds 
per 
site, respectively, 
constrained as $A \cup B = \{${\it all bonds}$\}$ and 
$A \cap B = \emptyset$. 
The ground-state degeneracy is equal to the number of the 
dimer-coverings.    
On a ladder and the higher dimensional lattices the number of 
the bond-coverings is infinity as the lattice size increasing 
to 
infinity. 
These ground states are kinds of the RVB states and the new 
types of 
the 
VB 
states induced by the interplay between the spin and orbital 
degrees 
of 
freedom. 
Taking some additional interaction terms to the Hamiltonian, 
one can restrict the number of the degenerate states.       
The ground state of the Hamiltonian: 
    ${\cal H} \to {\cal H} + x \sum_{(i,j) \in {\rm rungs}} 
     \left( {\cal P}_{i,j}^{(S=1)} - \frac{3}{4} \right) 
     + y \sum_{(i,j) \in {\rm legs}} 
     \left( {\cal P}_{i,j}^{(\tau= 2)} - \frac{1}{3} \right)$, 
$x,y >0$ is a product of all rung dimers in the spin sector 
and two decoupled VBS states in the orbital sector.
Therefore, the excitation energy has a finite gap\cite{AKLT}. 
Another Hamiltonian:
    ${\cal H} \to {\cal H} + x \sum_{m}
    {\cal H}_{\rm MG}^{(m)}$($x > 0$) 
where ${\cal H}^{(m)}_{\rm MG}$ denotes the Hamiltonian of 
the Majumder-Ghosh model on the chain 
$m\,(m=1,2)$\cite{Majumder-Ghosh}.
The ground states are four-fold degenerate and have two types. 
One is a checkerboard-type product of the singlet dimers along 
the 
legs 
in the spin sector 
and the VBS state connecting a chain in the orbital sector 
(fig.1(a)).
The other is a product of the two neighboring dimers along the 
legs 
in the spin sector and a product of the 4-site plaquette VB 
states 
in the orbital sector (fig.1(b)).
These are the localized states as the dimer, the VB, and the 
finite 
VBS
states.
In the case of above two Hamiltonians, it is expected that
each excitation energy has a finite gap.
    
In general, the model can be easily extended to the higher 
spin and 
pseudo-spin cases. 
We can construct the models with the spin-$S$ and the 
pseudo-spin-$T$ 
on the lattices whose coordination number is $z$, 
which have the following exact VB ground states:
  \begin{eqnarray} 
    | \Psi_{{\rm G}} (A,B) \rangle &=& 
    | \Phi^{\rm s} (A) \rangle \otimes | \Phi^{{\rm \tau}}(B) 
\rangle,
     \\
    | \Phi^{\rm s} (A) \rangle   &=& \prod _{(i,j) \in A} 
    ( a_i^{\dagger} b_j^{\dagger} - b_i^{\dagger} 
a_j^{\dagger} 
)^{M_s} 
    | 0 \rangle, 
    \nonumber \\
    | \Phi^{{\rm \tau}}(B) \rangle &=& \prod _{(i,j) \in B} 
     ( c_i^{\dagger} d_j^{\dagger} - d_i^{\dagger} 
c_j^{\dagger} 
)^{M_\tau}
      | 0 \rangle.
     \nonumber   
    \end{eqnarray} 
Here $M_s$ and $M_\tau$ obey $M_s = 2S/z_s = {\rm integer}$ 
and 
$M_\tau = 
2T/z_\tau = {\rm integer}$, respectively, 
and $z$ satisfies $z \le z_s + z_\tau$. 
$A(B)$ denotes a configuration of the bonds which has 
$z_s(z_{\tau})$ 
bonds 
per site and $A \cup B =\{$ {\it all bonds} $\}$.
The Hamiltonian with these ground states is given by
   \begin{eqnarray}
   {\cal H} = \sum_{(i,j)} \sum_{J=J_0}^{2S} \sum_{I=I_0}^{2T}
   K_{J,I} {\cal P}_{i,j}^{({\bf S} = J)}{\cal P}_{i,j}^{({\bf 
\tau} 
= 
I)},
    \end{eqnarray}
where $J_0=2S-M_s+1$, $I_0=2T-M_{\tau}+1$, and $K_{J,I} \ge 
0$. 
The simplest example of the one-dimensional models is given by 
    ${\cal H} = \sum_{(i,j)} [{\bf S}_{i} \cdot {\bf S}_{j} + 
S(S+1) 
]
    [ {\bf \tau}_{i} \cdot {\bf \tau}_{j} + T(T+1) ]$, 
whose ground states are the singlet dimer states with two-fold 
degeneracy 
in both the spin and the pseudo-spin spaces, 
where $(z,z_s,z_\tau) = (2,1,1)$. 
Moreover, one can easily extend to the anisotropic cases. 
It is natural that the SU(2) symmetry is broken in the orbital 
sector. 
The models have the same ground states of the isotropic cases 
as long as the lowest level of the two-site local Hamiltonian 
does 
not 
cross the higher levels.
    
    In summary, we have proposed a method to obtain 
the exact ground states of the models 
which are composed of the products of two commutable 
operators. 
Applying this method to the quantum spin models with the 
orbital 
degeneracy as the Hamiltonian (\ref{eqn:Ham1}), 
we have obtained various ground states of such models and 
found that these models have the following interesting 
properties 
under particular parameter regimes.  
The frustration arising from the spin-orbital interaction 
of the transverse direction 
increases the quantum fluctuation 
and induces the singlet dimer ground states. 
The long range orders in one sector enhance various 
instabilities in 
the 
other sector in any dimension, 
e.g., the orbital orders make the systems have the spin liquid 
states 
and the AFLRO exist with infinitesimal anisotropy of the 
$z$-direction
on the bipartite lattices.
In addition, we have constructed models with arbitrary spin 
and 
orbital degeneracy which have the exact VB ground states 
whose degeneracy is infinity in the thermodynamic limit on a 
ladder 
and 
higher dimensional lattices. 
These VB states include the correlations between the spin and 
orbital 
degrees of freedom and are the new types of the RVB states in 
D $> 
1$.  
The exact ground states are obtained under restrict parameter 
regions, but we expect that these states are 
adiabatically connected with the ground states in the large 
parameter 
space. 

The author would like to thank M. Takahashi, T. Kawarabayashi,
M. Nakamura and N. Muramoto for their encouragement and 
helpful
comments.

\begin{figure}
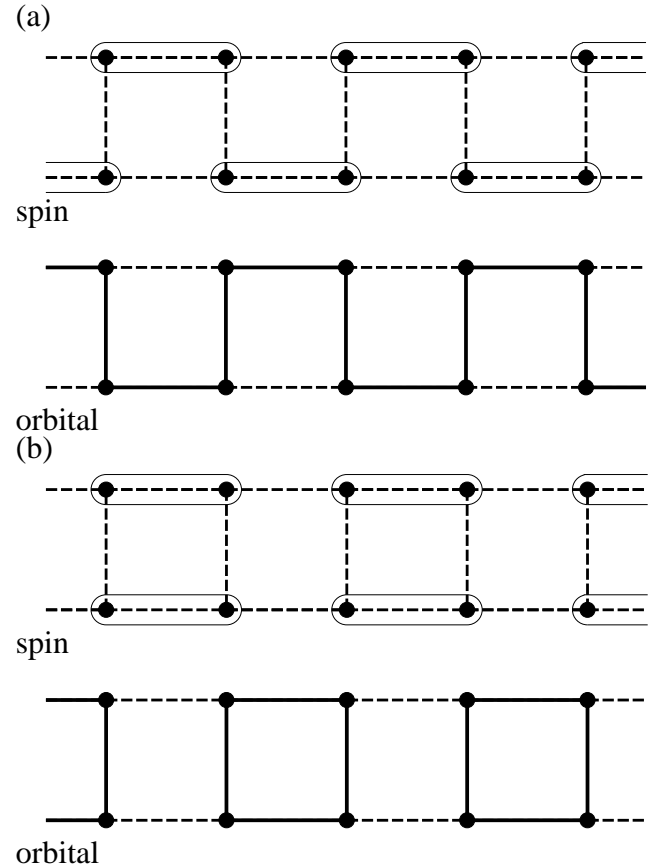

 \epsfxsize=3.0in \leavevmode \epsfbox{ladder1.eps}\\

 \epsfxsize=3.0in \leavevmode \epsfbox{ladder2.eps}\\

\caption{Two types of the VB states on the ladder lattice (see
 text). Pairs in the spin sector indicate dimer states.
Solid lines in the orbital sector denote (a) VBS, (b) finite 
VBS 
states.
 \label{fig:ladders}}
\end{figure}

\end{document}